\documentclass[10pt,twocolumn,twoside]{IEEEtran}

\usepackage{amsfonts,amssymb,amsbsy,amsmath,paralist,theorem,bm,ifthen,color}

\usepackage{graphicx}
\usepackage{subfigure,relsize}

\usepackage[noadjust]{cite}
\usepackage[numbers,sort&compress,square]{natbib}

\usepackage{cases,booktabs}
\usepackage{flushend}

\usepackage{algorithm}
\usepackage{algpseudocode}

\usepackage{cuted}
\usepackage[dvipsnames]{xcolor}

\usepackage{multirow}
\usepackage{makecell}
\usepackage{array}

\usepackage{multirow}

\definecolor{applegreen}{rgb}{0.55, 0.71, 0.0}

\definecolor{mycyan}{HTML}{4b5cc4}

\newcommand\Mc{\ensuremath{{\mathcal{M}}}}

\newcommand\xb{\ensuremath{{\bm x}}}

\newcommand\Hb{\ensuremath{{\bm H}}}

\newcommand\Tb{\ensuremath{{\bm T}}}

\newcommand\zb{\ensuremath{{\bm z}}}

\graphicspath{{figures/}}

\begin{document}

\title{PEMNet: Towards Autonomous and Enhanced Environment-Aware Mobile Networks}

\author{Lei Li, 
Yanqing Xu, Ye Xue, Feng Yin, Chao Shen, Rui Zhang, and Tsung-Hui Chang

\thanks{\smaller[1] Lei Li, Yanqing Xu, Feng Yin, Chao Shen, and Tsung-Hui~Chang are with the Chinese University of Hong Kong, Shenzhen (CUHK-Shenzhen) and also with the Shenzhen Research Institute of Big Data (SRIBD), China. Ye Xue is with the Sun Yat-sen University (SYSU), China. Rui Zhang is with the National University of Singapore (NUS) and CUHK-Shenzhen.}

}

\maketitle

\begin{abstract}		

With 5G deployment and the evolution toward 6G, mobile networks must make decisions in highly dynamic environments under strict latency, energy, and spectrum constraints.  Achieving this goal, however, depends on prior knowledge of spatial-temporal variations in wireless channels and traffic demands.  This motivates a joint, site-specific representation of radio propagation and user demand that is queryable at low online overhead. In this work, we propose the perception embedding map (PEM), a localized framework that embeds fine-grained channel statistics together with grid-level spatial-temporal traffic patterns over a base station’s coverage. PEM is built from standard-compliant measurements—such as measurement report and scheduling/quality-of-service logs—so it can be deployed and maintained at scale with low cost. Integrated into PEM, this joint knowledge supports enhanced environment-aware optimization across PHY, MAC, and network layers while substantially reducing training overhead and signaling. Compared with existing site-specific channel maps and digital-twin replicas, PEM distinctively emphasizes (i) joint channel–traffic embedding, which is essential for network optimization, and (ii) practical construction using standard measurements, enabling network autonomy while striking a favorable fidelity–cost balance.

\end{abstract}

\section{Introduction} \label{sec:intro}

\subsection{Background and Motivations}

As 5G matures and 6G research accelerates, mobile networks are expected to support unprecedented levels of connectivity, capacity, and service diversity. These networks must make real-time decisions under stringent latency, energy, and spectrum constraints, while adapting to increasingly complex and dynamic environments. However, in practical scenarios, the network performance is dominated by two factors: (i) wireless propagation,  captured by signal propagation through complex environments, and (ii) user traffic, characterized by heterogeneous demand patterns that vary across time and location. 
{Specifically, in multi-input multi-output (MIMO) systems, periodic channel state information (CSI) acquisition is essential to track the time-varying propagation environment. However, this process incurs significant overhead that scales with both the number of antennas and the number of users. In multi-cell coordinated transmission, the challenge extends beyond acquiring CSI across cells—aggregating and sharing this information for joint scheduling and transmission optimization further consumes substantial spectrum and backhaul resources, limiting scalability and efficiency. Moreover, conventional designs are ill-adapted to diverse traffic types and often allocate resources based on coarse, static metrics rather than dynamic demand. This results in poor provisioning for traffic hotspots and inefficient resource use, highlighting the need for demand-aware, adaptive network management. These bottlenecks are further intensified by the rapid growth of wireless devices, bandwidth-intensive applications and increasingly diverse services.

In recent years, digital twin (DT) for mobile networks has drawn great research attention. With the ambitious goal to create a virtual replica of the physical mobile network that is continuously synchronized through real-time data to reflect its state and behavior, DT is envisioned to simulate, predict, and optimize network performance and resource management in a safe, data-driven, and proactive manner. However, existing DT systems are typically restricted to offline datasets of application traffic and device trajectories to learn and generate individual user behaviors, while employing ray-tracing for wireless signal propagation simulation \cite{DT_sustainable_mag25}. Such modeling falls short of achieving the environment perception continuously synchronized with the live network, thereby confining current DT systems to offline algorithm evaluation and what-if analysis. To promote digital twin channel (DTC), radio environment knowledge pools (REKPs) \cite{REKP_mag25} were proposed to characterize environment features related to channels, while channel knowledge map (CKM) \cite{CKM_tut24} introduced site-specific databases providing location-specific channel knowledge. Though these ideas are inspiring, they overlook user data traffic perception. Additionally, most existing data traffic prediction paradigms operate at the cell or network level \cite{Traffic_JSAC19}, failing to capture fine-grained spatial distribution and thus limiting their application in sophisticated network optimization.

\begin{table*}[t]
	\centering
	\caption{Comparison of PEM and other existing fine-grained wireless environment maps.}
	\begin{tabular}{|c|c|c|c|c|}
		\hline
		\textbf{Technique} & \textbf{User Data Traffic} & \textbf{Channel Statistics} & \textbf{Channel Measurements} & \textbf{Focus} \\ \hline
		\textbf{\begin{tabular}[c]{@{}c@{}}Channel charting\\ (CC) \cite{CC_access18}\end{tabular}} & Not provided & Large-scale & CIR & \begin{tabular}[c]{@{}c@{}}Localize users from  \\ channel data \end{tabular}\\ \hline
		\textbf{\begin{tabular}[c]{@{}c@{}}Channel knowledge \\ map (CKM) \cite{CKM_tut24}\end{tabular}} & Not provided & \begin{tabular}[c]{@{}c@{}}Large-scale and/or \\ small-scale \end{tabular} & \begin{tabular}[c]{@{}c@{}}Channel data \\ (dependent on knowledge type)\end{tabular} & \begin{tabular}[c]{@{}c@{}}Provide location-specific  \\  channel knowledge\end{tabular} \\ \hline
		\textbf{\begin{tabular}[c]{@{}c@{}}Radio environment \\ knowledge \\ pool (REKP) \cite{REKP_mag25} \end{tabular}} & Not provided & \begin{tabular}[c]{@{}c@{}}Large-scale and \\ small-scale\end{tabular} & CIR & \begin{tabular}[c]{@{}c@{}}Provide channel-related \\ environment features  \end{tabular}  \\ \hline
		\textbf{\begin{tabular}[c]{@{}c@{}}Perception \\ embedding \\ map (PEM)\end{tabular}} & Provided & Large-scale & Multi-beam RSRP & \begin{tabular}[c]{@{}c@{}} Embed localized channel \& traffic \\ knowledge from practical \\   standard measurements\end{tabular}  \\ \hline 
	\end{tabular}
	\label{Tb:comparsion}
	\vspace{-0.1cm}
\end{table*}

To address the aforementioned issues, mobile networks require fine-grained perception of both wireless propagation and data traffic, with these two knowledge dimensions being exploited synergistically. For instance, in a multi-cell network where each base station (BS) has no explicit knowledge of active users in other cells, a fine-grained traffic spatial map can enable the BS to infer traffic hotspots, which are integrated with a fine-grained channel map to estimate inter-cell interference, thereby guiding transmission design to improve cell edge performance. Neither a channel map nor a traffic map working in isolation can adequately address this problem. Further, to support autonomous network optimization, environmental perception data must be easily measurable from practical systems in a cost-efficient manner. Yet, this remains largely unaddressed in existing paradigms, which assume ideal access to user trajectories, application traffic, and channel impulse responses (CIRs), which are not always available in practice due to privacy and system constraints. Acquiring such data requires dedicated sensing equipment and measurement campaigns, which increase costs and hinder practical deployment.

These practical demands for autonomous, environment-aware mobile networks motivate the development of a more comprehensive and data-efficient environment perception paradigm. To this end, we propose the {\bf perception embedding map (PEM)}: a site-specific, queryable database that maps space–time features (STFs) to grid-level channel knowledge and traffic demand, constructed from standard measurements and designed for frequent, low-overhead queries in large-scale live networks. Specifically, PEM owns the following features.
\begin{enumerate}
	\item {Multi-dimensional perception}: providing both channel knowledge and user data traffic knowledge. On the one hand, by leveraging a CKM that provides large-scale features such as the angular power spectrum (APS) and the delay power spectrum (DPS) of dominant paths, the channel perception in PEM captures the key statistical characteristics of the signal propagation through multi-path environments. 
	{On the other hand, the user data traffic perception in PEM captures fine-grained spatial-temporal patterns of user demands and activity -- such as where and when users are active, their demand forecasts, and alignment with link reliability.
	This enables accounting for demand-driven factors like inter-cell interference, MAC-layer physical resource block (PRB) allocation, and network-level handover parameters. By integrating both wireless propagation and user traffic patterns, PEM enhances environment perception and facilitates joint optimization across layers. }	
	\item{Low-cost and better applicability}: PEM is constructed using measurement mechanisms standardized in commercial cellular networks such as measurement
	report (MR), avoiding the need for dedicated sensing hardware equipment and large-scale measurement campaigns. For channel perception, PEM estimates {statistical} characteristics from reference signal received power (RSRP) measurements, rather than relying on full CIR, which is more complicated, resource-demanding, and measurement-limited. 
	For traffic perception, PEM leverages historical traffic data statistics, user equipment (UE) connection logs, and scheduling records that are well defined by 3GPP and easy to access in current 4G/5G networks. While creating significant challenges in environment perception, these accessible and standardized data sources make PEM cost-efficient, scalable, and directly applicable in real-world deployments.
\end{enumerate}
}

\subsection{Existing Paradigms}

\begin{figure*}[htbp]
\centering
\includegraphics[width=1.0\textwidth]{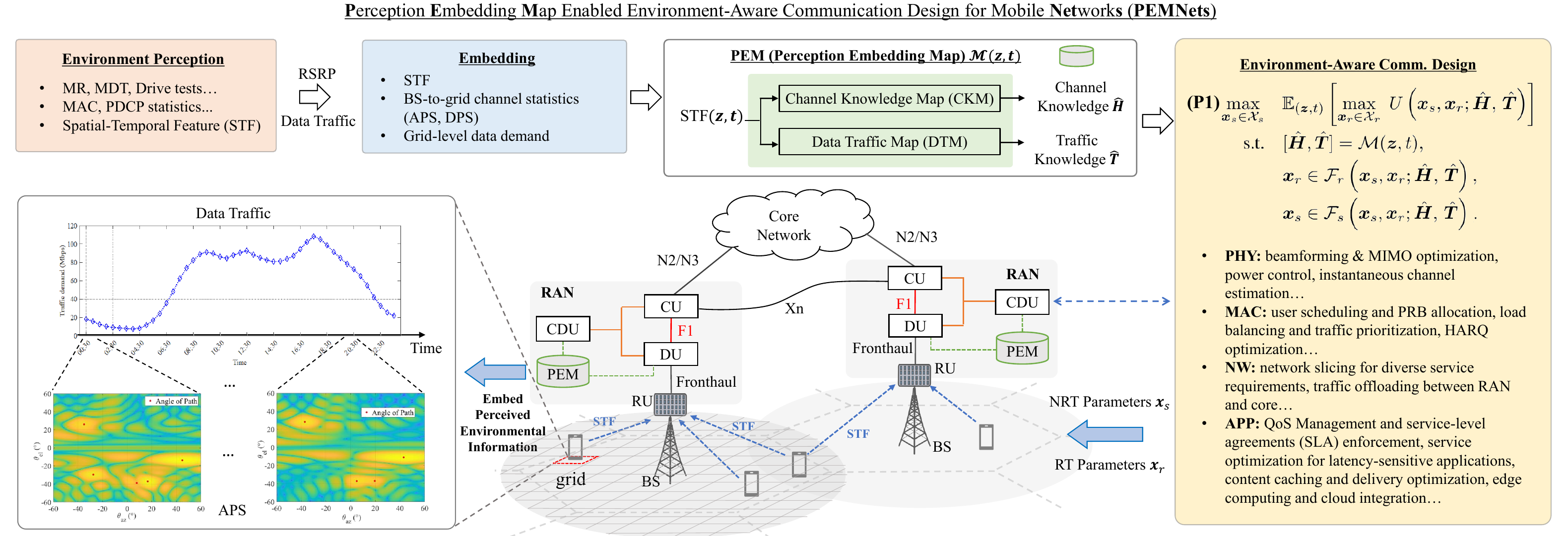} 
\caption{The PEM is a site-specific database that embeds the perceived environment information to provide statistical channel knowledge $\hat\Hb$ and the predicted data traffic demand $\hat\Tb$. The overall coverage area of the site (BS) is divided into multiple spatial grids. With the STFs $\{(\zb, t)\}$ of UEs as inputs, the PEM outputs their corresponding $\{(\hat\Hb, \hat\Tb)\}$ to support the environment-aware communication design of mobile networks. In the PEM-enabled open radio access network (O-RAN), each RAN consists of one or several radio units (RUs), a distributed unit (DU), a centralized unit (CU), a communication design unit (CDU) and a PEM. The PEM is connected to DU for data collection and STF inputs, and its output is fed into the CDU to conduct environment-aware optimization.   }  
\label{fig:PEMNet}
\vspace{-0.3cm}
\end{figure*}

{Conventional radio maps} mainly refer to databases of geographical signal power spectrum density or large-scale channel gains like the one-dimension path loss used for coverage planning and mobility management \cite{ERM_WC19}. 
Due to their provision of only coarse-grained channel statistics and lack of detailed propagation path information, their applicability in multi-antenna networks is limited.
Additionally, though {user data traffic prediction} in wireless networks has been extensively investigated, most works focus on aggregated traffic at the BS or cell level  \cite{Traffic_JSAC19}. While such approaches are suitable for coarse-grained resource management like BS sleep control, their spatial resolution is insufficient to capture traffic variations across a smaller spatial scale, for instance, intra-cell grid locations at meter-scale granularity required for fine-grained communication design. 

{ In recent years, the pressing need for enhanced environmental awareness in intelligent mobile networks has spurred a series of representative research efforts on fine-grained wireless environment maps, such as channel charting (CC) \cite{CC_access18}, CKM, and REKP. }
Specifically, CC constructs low-dimensional channel features that capture the local spatial geometry of users from channel measurements, thereby enabling relative user localization and mobility inference for tasks such as user tracking and handover. In contrast, by mining channel measurements tagged with locations, {CKM} maps user locations to their channel knowledge -- a broad concept that ranges from coarse channel gain to complete CIR. Additionally, REKP aims to extract channel-related environment features such as scatter and blockage characteristics to facilitate DTC.

{Table \ref{Tb:comparsion} provides a detailed comparison of these studies with our proposed PEM.} 
First, unlike CC, CKM and REKP, PEM aims to build a grid-level spatial-temporal map, where user data traffic knowledge together with channel knowledge are embedded in each grid, reflecting the important design consideration in modern network optimization that effective communication design benefits from understanding both the radio propagation environment and user traffic patterns.  Second, PEM highlights the fidelity-cost balance in practice, featured by the large-scale channel knowledge it provides and its utilization of standardized measurements. The CIR measurements required by REKP, CC, and CKMs for small-scale parameters incur substantial channel estimation (CE) overhead and large memory demands, and are constrained in the downlink of frequency duplex division systems and inter-cell channels, since only compressed/quantized measurements such as channel quality indicator (CQI) or RSRP are fed back to BSs, respectively. This poses great challenges to data measurements, storage, and computation.  In contrast, PEM's RSRP-based channel perception leverages readily available standardized mechanisms like MR to avoid these issues, whereas its extracted channel knowledge still captures essential multipath characteristics. This practical deployment consideration is also featured by PEM's grid-level traffic perception by utilizing scheduling logs and quality-of-service (QoS) statistics readily available at the BS rather than from fine-grained user data (e.g., trajectory, application traffic type and demand) \cite{DT_sustainable_mag25} that is highly restricted at BSs. These features collectively enable PEM to achieve better operational practicality while preserving key environment knowledge.

\section{PEM: Environment Perception \& Embedding} \label{sec:model}

As illustrated in Fig. \ref{fig:PEMNet}, PEM is a site-specific database deployed in a BS to capture the localized environment knowledge. The coverage area of a BS is divided into multiple spatial grids. The PEM comprises a CKM to provide the time-evolving large-scale statistics of the propagation channel $\hat \Hb$, and a data traffic map (DTM) that offers the data traffic knowledge $\hat \Tb$ at each grid over time. Specifically, the large-scale channel knowledge includes the APS and the DPS of dominant paths of the channel between the BS to each grid, and the traffic knowledge refers to the user data traffic demands at each grid. Given STFs $(\zb, t)$ with $\zb$ being the spatial feature (e.g., location coordinates or grid indices) and $t$ being the temporal feature (e.g., timestamp), PEM returns the corresponding environment knowledge as a mapping $[\hat\Hb, \hat\Tb] =\Mc(\zb, t)$. 
Here, the granularity of STFs depends on the correlation of knowledge provided across space and time, and its configuration needs to balance the trade-off between accuracy and processing complexity. For example, as the large-scale channel characteristics are determined by stable environmental factors (e.g., static scatterers like buildings, terrain, and large obstacles)  and the transmitter-receiver geometry, they not only exhibit strong spatial consistency but also evolve slowly over time. Therefore, the spatial granularity of PEM (e.g., grid size) can be set to several meters and the time granularity can be dozens of minutes to hours.

To establish $\Mc(\cdot)$, PEM construction involves two stages: an {\bf environment perception} stage to collect measurement data, and an {\bf  information-embedding} stage that builds models for extracting the STF, BS-to-grid channel statistics and grid-level data demand from these measurements.

\subsection{Environment Perception}
The PEM construction begins with collecting diverse wireless network measurements tagged with STFs. A distinct feature of PEM is the data that can be gathered by the mature measuring and reporting mechanisms standardized in practical cellular networks. This avoids the need for specialized measurement equipment, extensive manual campaigns, and significant modifications to existing network infrastructure. 

Specifically, the RSRP measurements used for CKM construction can be obtained through the following approaches: {\bf a) MR } \cite{MR_TS36331} leverages user devices to passively report RSRP, reference signal received quality (RSRQ), and serving/neighboring cell IDs during routine events (e.g., handovers, radio resource control measurements). MR provides broad coverage but may suffer from irregular spatial distribution due to user mobility and event-triggered reporting, while also lacking UE location coordinates \cite{Xinyu_ArXiv25}. {\bf b) Minimization of drive test (MDT) } introduced by 3GPP [2] complements MR by enabling automated, network-controlled (immediate MDT) or UE-initiated (logged MDT) data collection. Compared with MR, MDT reports richer information, such as UE location and throughput statistics, but it may not always be implemented at all UEs due to privacy and power constraints. {\bf c) Small-scale drive test} involves driving a vehicle equipped with specialized equipment along predefined routes to actively collect and record accurate, high-resolution network performance data, including RSRP, signal-to-interference-plus-noise-ratio (SINR), and global positioning system (GPS) coordinates. While being precise, this method requires dedicated equipment and measurement campaigns, resulting in higher costs and limited scalability.
In propagation environment perception, these approaches are complementary. 

The merits of automation, low-cost, and wide-area coverage make MR and MDT ideal for continuous, large-scale, crowdsourced measurements, forming the basis for autonomous environment perception. In addition, small-scale drive tests are suitable for generating high-quality reference datasets in regions with complex propagation conditions (e.g., urban canyons and tunnels) or areas requiring precise calibration. Their integration ensures both accuracy and scalability in propagation perception.

Meanwhile, the {user data traffic} for DTM construction can be measured at the BS through the control and user-plane monitoring mechanisms \cite{MDT_TS23501} of current mobile networks, elaborated as follows: {\bf a) MAC-layer statistics} provide real-time traffic load, including the number of active UEs, allocated PRBs, scheduled throughput, and buffer occupancy; {\bf b) Packet data convergence protocol (PDCP) layer statistics} record higher-layer traffic characteristics such as transmitted/received data volume, packet arrival rate, retransmissions, and delay. Compared with the MAC-layer data, PDCP statistics capture more stable user-plane behavior over longer timescales; {\bf c) MDT} also reports traffic statistics, including throughput, delay and packet loss, along with UE location. By integrating MAC/PDCP statistics and MDT reports with user locations or beam indices derived from channel measurements, the BS can learn grid-level traffic patterns across space and time.

\subsection{Information Embedding}

{\bf STF Embedding}: The core task herein is to extract spatial features $\zb$ from measurement data. Unlike temporal features (e.g., timestamps), UE location information -- though being available at the device (e.g., via GPS) -- is typically inaccessible to the BS in MR measurements. While drive tests and MDT reports can provide accurate locations, their coverage is less extensive. Hence, STF embedding must infer $\zb$ from accessible data without relying on geographical coordinates. This is crucial for the deployment in practice. One approach is to construct $\zb$ as a virtual spatial feature in the beam/channel space, which can be discretized into grids such that each grid shares consistent channel structures. As the beam/channel space feature is inherently determined by the user’s physical location and the surrounding propagation environment,  the virtual features implicitly encode the user's spatial signatures relative to the BS. Importantly, these features can be derived from readily available metrics such as multi-beam RSRP, which aggregates multipath power contributions across spatial directions. Recent advances, including grid construction \cite{Xinyu_ArXiv25} and channel space gridization \cite{Juntao_ArXiv25}, have demonstrated promising performance of STF embedding from MR measurements.

{\bf Channel Embedding}: As a key module of PEM, the CKM can be constructed in a non-parametric manner. Specifically, given RSRP samples with STFs, the BS-to-grid APS and DPS are extracted and stored for different temporal intervals. According to localized statistical channel modeling (LSCM) \cite{LSCM_twc23}, a linear relationship exists between multi-beam RSRP measurements and their corresponding channel APS, enabling APS recovery to be formulated as a sparse signal reconstruction problem. The main challenge in recovery lies in the ill-conditioned coefficient matrix of the problem, which degrades estimation accuracy. To mitigate this, high-order RSRP statistics can be exploited to enhance recovery robustness \cite{LSCM_twc23}, while leveraging multi-grid structured sparsity and spatial consistency across adjacent grids can further improve estimation. Moreover, by capturing the interactions between the environment and radio waves through neural radiance fields (NeRF$^2$) modeling, the physics-informed learning approach \cite{Binsheng_ArXiv25} can achieve improved recovery performance across multi-cell, multi-grid, and multi-frequency measurements.

For grids without direct RSRP measurements, the large-scale channel parameters can be inferred from neighboring grids by leveraging their inherent spatial consistency through classical spatial interpolation techniques such as Kriging and kernel regression. 
Alternatively, by modeling the grids as a graph -- where each node represents a grid and edges encode the spatial similarity of features -- while treating APS/DPS as graph signals, graph signal processing models can be employed for spatial-domain interpolation. Besides, data-driven interpolation approaches such as 3D Gaussian splatting (3DGS) that have demonstrated strong capabilities in capturing complex spatial dependencies can be utilized. 
Together, these methods enhance the channel knowledge recovery across regions with missing measurements, thereby ensuring the completeness and spatial continuity of the CKM.

{\bf Traffic Embedding}: The DTM construction will build a mapping that predicts the traffic volume at each grid over time, in contrast to conventional cell-level prediction. Given that traffic patterns often exhibit strong temporal regularities -- such as daily or weekly cycles -- the DTM can be established by learning structured predictive models that map temporal features (e.g., time-of-day, day-of-week) to grid-level traffic volume \cite{Traffic_JSAC19}. This modeling not only enables generalization to unseen time instances but also enhances robustness against small data with outliers and non-Gaussian measurement noise, as the imposed structural assumptions, such as smoothness, global and local periodicity, and structured temporal dynamics, serve as inductive biases to filter out noise and prevent overfitting to outliers.

To learn such grid-level spatial-temporal traffic patterns, various techniques can be adopted \cite{traffic_acmSurvey25}. Non-parametric Bayesian frameworks such as Gaussian process regression (GPR), which model traffic as a distribution over functions, can provide both point prediction and uncertainty quantification. By incorporating temporal and spatial kernels, GPR can capture smooth trends and periodic variations, making it particularly suitable for small data with Gaussian noise. 
Spatial-temporal neural networks, such as ConvLSTM extending recurrent architectures by integrating convolutional operations, allow the model to simultaneously learn localized spatial patterns and temporal dependencies, while embedding both short-term fluctuations and long-term periodic trends in traffic dynamics.
Moreover, graph neural networks are superior in modeling traffic correlations across grids, improving inference quality in sparsely sampled environments.

\section{PEMNet: PEM-enabled optimization for mobile networks  }  \label{sec:CQI_Q}

Once constructed, the PEM delivers fine-grained spatial-temporal channel and traffic knowledge in the coverage area of each site via $[\hat\Hb, \hat\Tb] =\Mc(\zb, t)$ (see Fig. \ref{fig:PEMNet}), thereby enabling intelligent optimization of diverse communication tasks of mobile networks. To characterize the interaction among environment knowledge, optimization decisions, and performance objectives, we unify the PEM-enabled communication optimization as problem (P1) in Fig. \ref{fig:PEMNet},
where $U(\cdot)$ is a generic utility function, such as the weighted sum-rate, energy efficiency, or a channel acquisition metric, and $\xb_s$ and $\xb_r$ are the long-term/non-real-time (NRT) variables and the short-term/real-time (RT) variables to be designed, respectively.  $\mathcal{F}_s(\cdot)$ and $\mathcal{F}_r(\cdot)$ are the feasibility sets defined by system and physical constraints (e.g., QoS, capacity, power budget), dependent on both the PEM outputs and the optimization variables.

In this framework, PEM captures the user distribution and traffic dynamics by the DTM and characterizes large-scale channel features between the BS and grids via the CKM, offering a comprehensive and fine-grained view of the network environment.
The knowledge output by PEM can be directly applied to NRT communication design and proactive resource allocation, such as BS placement, antenna configuration, cell splitting, and spectrum reuse strategies \cite{SRCON23}. Further, based on the outputs, BSs can perform lightweight measurements to acquire instantaneous environmental states, enabling RT communication optimization tasks.  Accordingly, PEMNet enables more efficient transmission designs in different time scales and protocol layers.

\section{Case Studies}

In this section, we present two concrete examples to demonstrate how PEMNet can improve wireless system designs. 

\subsection{PEMNet-enabled Multi-Cell Coordinated Beamforming}

\begin{figure}[t]
\centering
\includegraphics[width=0.48\textwidth]{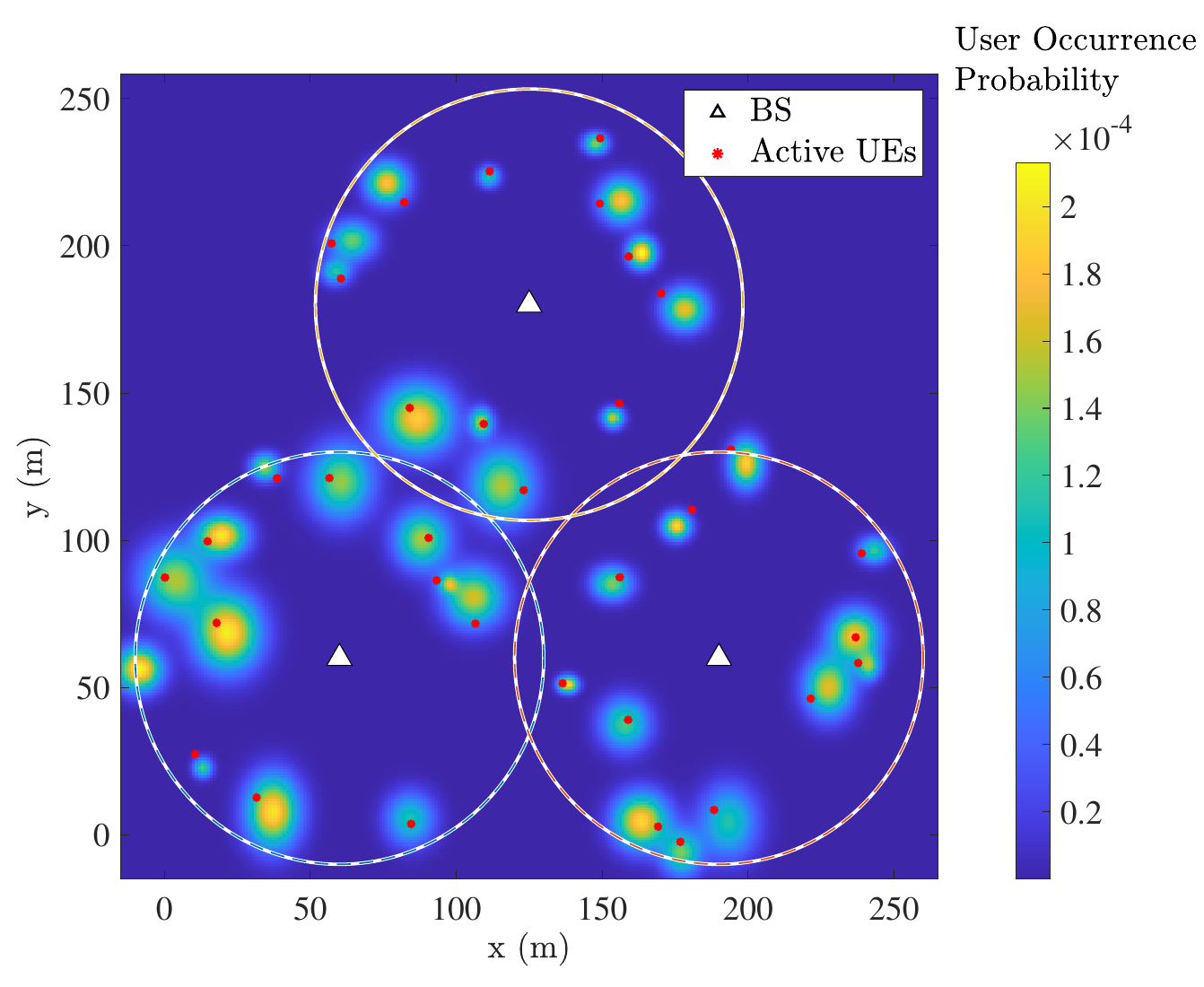}  
\caption{A multi-cell network including 3 BSs and 36 UEs.   
}
\label{fig:topo}
\end{figure}

\begin{figure}[t]
\centering
\includegraphics[width=0.48\textwidth]{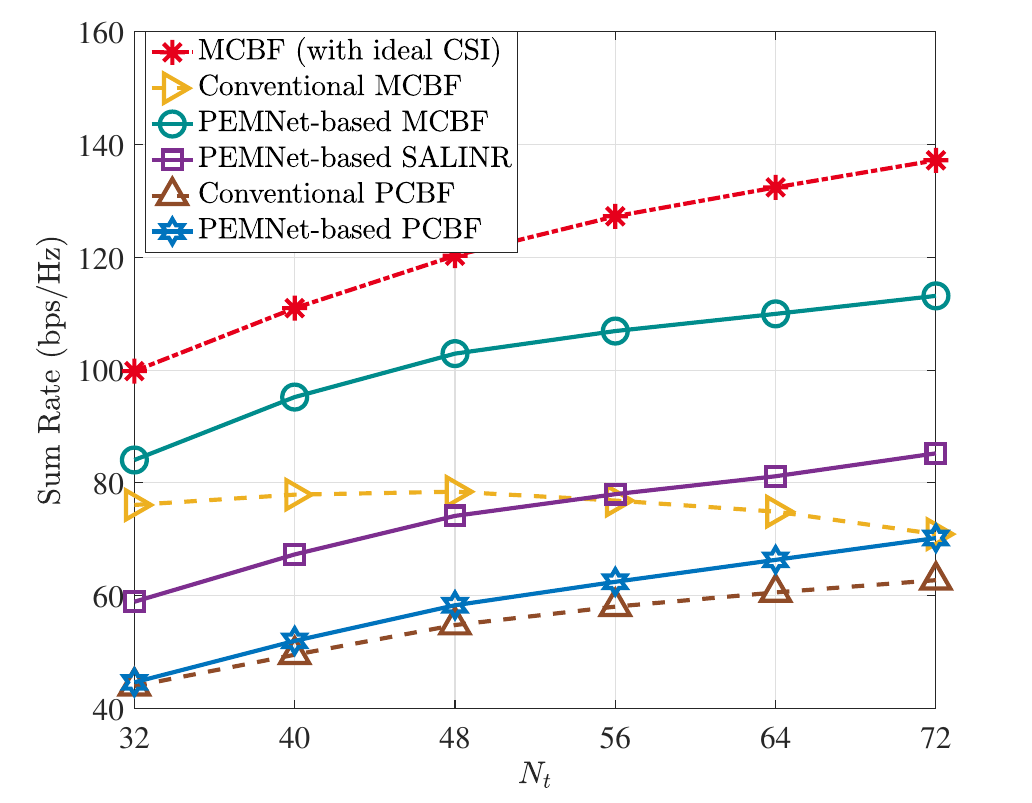}  
\caption{ESR achieved by different BF schemes versus $N_t$, $n_0 = 500$.   
}
\label{fig:ESR_MCBF}
\end{figure}

Multi-cell coordinated beamforming (MCBF) has been widely recognized as a promising way to enhance the performance of mobile networks, especially for cell-edge users. However, conventional MCBF needs to take considerable CE overhead to acquire intra-cell and inter-cell CSI, which reduces the effective network throughput. The proposed PEMNet overcomes these limitations by providing BSs with enhanced environment awareness. As shown in Fig. \ref{fig:topo}, we evaluate conventional MCBF and PEMNet-enabled MCBF in a network with $36$ single-antenna UEs and $3$ cells, each served by a BS with $N_t$ antennas. The heatmap demonstrates the user occurrence probability inferred from the DTM of PEM. The downlink channels from BSs to UEs comprise $n_p=4$ propagation paths with random angles and phases. In this illustrated example, the APS of each grid is extracted from RSRP measurements, and the data traffic across grids is generated by a Gaussian mixture model. In PEMNet-based schemes, each BS first retrieves the APS from its CKM, and only takes a slight overhead to estimate the instantaneous channel gains to obtain the  CSI. 
In Fig. \ref{fig:ESR_MCBF}, except the curve of MCBF (with ideal CSI) that represents the sum rate without deducing the CE overhead, all the other results correspond to the effective sum rate $\text{ESR} \triangleq (n_0-n_1)/n_0\sum_{k=1}^{K}\text{log}_2(1+\gamma_k)$, where $n_0$, $n_1$, and $\gamma_k$ denote the total number of symbols in the coherence period, the number of pilot symbols, and the SINR of UE $k$, respectively. All MCBF schemes employ the weighted minimum mean square error (WMMSE) algorithm for sum-rate maximization, and per-cell beamforming (PCBF) without coordination is also included for comparison.  Comparing PEMNet-based MCBF (and PEMNet-based PCBF) with their conventional counterparts, the former achieves higher ESR owing to the overhead reduction enabled by the PEM. 

Additionally, PEMNet can still be effective even when messages cannot be exchanged between BSs. In this scenario, each BS can still infer the spatial distribution of the active users in its neighboring cells from its DTM and query its CKM to estimate the cross-cell leakage CSI by simply summarizing the associated APS. The integration of DTM and CKM narrows inter-cell leakage estimation to a sparse set of likely active grids, leading to a distributed MCBF solved by the SALINR algorithm \cite{Tenghao_ICASSP24} and termed as PEMNet-based SALINR. One can observe that PEMNet-based SALINR still largely outperforms the PCBF schemes even without explicit knowledge of active users in other cells, thanks to the dual-view knowledge offered by the PEM.

\subsection{PEMNet-assisted Receive Beam Design}

\begin{figure}[t]
\centering
\includegraphics[width=0.48\textwidth]{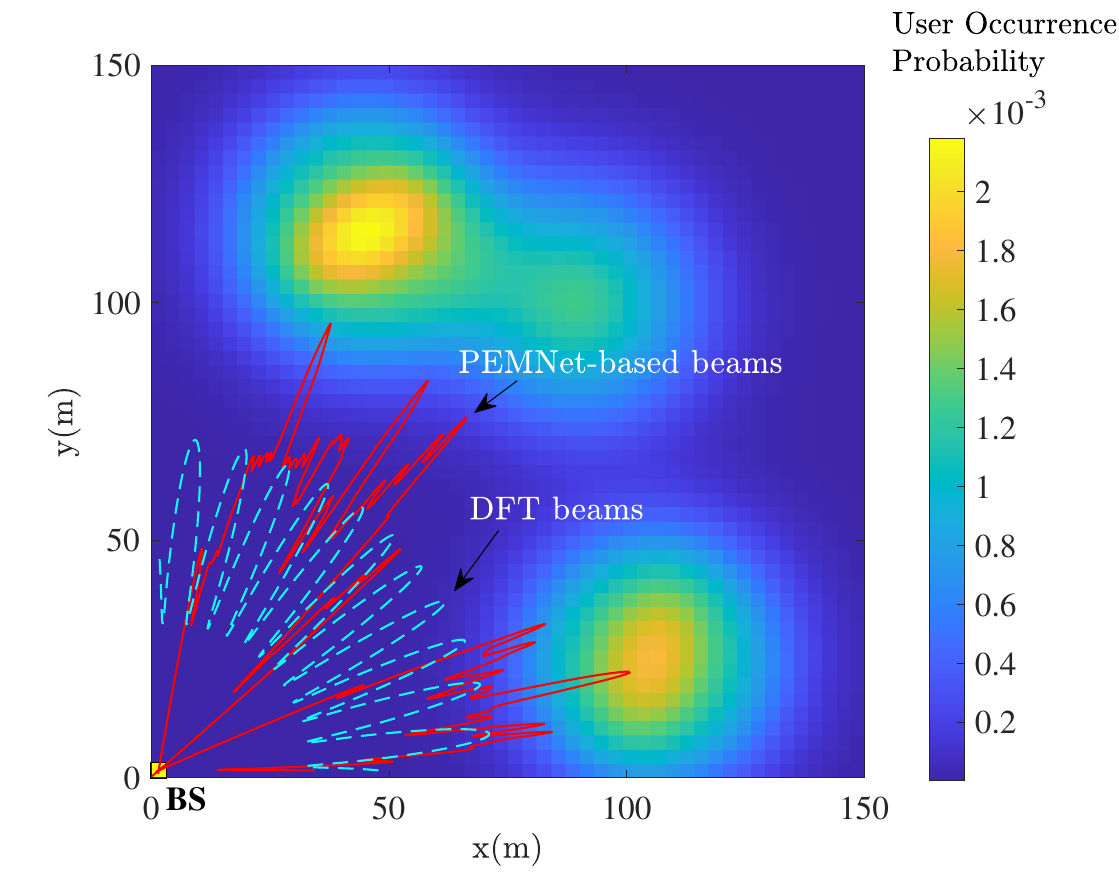}  
\caption{PEMNet-based beam intensity map aligned with user distribution hotspots.}
\label{fig:rx_bf}
\end{figure}

To effectively capture user signals across the coverage area, the BS typically adopts a discrete Fourier transform (DFT)-based receive matrix to span the entire angular domain. However, such a uniform design fails to exploit the spatially non-uniform user distribution. To address this limitation, we propose a PEMNet-assisted beamspace design. As shown in Fig. \ref{fig:rx_bf}, the service area is partitioned into fine spatial grids, each characterized by its user occurrence probability inferred from DTM and its APS from the CKM. By integrating both kinds of knowledge from the PEM, the BS reconstructs the local channel covariance from the APS of each grid and forms a weighted covariance matrix by aggregating these local covariances according to user occurrence probabilities. The dominant eigenmodes of this weighted covariance determine the optimal receive beamspace, which naturally aligns with user hotspots. In the illustrated example, a BS is deployed at the lower-left corner of a $150~\text{m} \times 150~\text{m}$ region containing three dense user clusters. The grid color indicates the user occurrence probability, while the red/blue contours represent the receive beam energy. Compared with conventional DFT beams, the proposed PEMNet-based design concentrates energy toward active regions, resulting in more power being focused on user hotspots and less leakage to low-demand areas.

\section{Extensions}

\subsection{Enhancing PEM Construction and Adaptation}
To further unlock the potential of PEM, several key aspects merit further exploration.   {\bf 1) Multi-Modal Data Fusion for Channel Perception:} 
Though the channel perception of PEM mainly takes readily accessible network measurements like RSRP, when additional sensory data from LiDAR, cameras, and radar is available, fusing these multi-modal inputs can further improve the accuracy of the extracted knowledge. To exploit multi-modal data, techniques like representation learning and latent space alignment can be utilized to unify heterogeneous features, while attention mechanisms adaptively weight modalities according to context (e.g., visual cues in non-line-of-sight or LiDAR geometry in urban canyons). These techniques collectively enhance the accuracy and resilience of radio propagation perception.
{\bf 2) Collaborative DTM Construction:} 
In practice, the traffic data at individual grids could be sparse, which motivates cross-site collaborative construction to enhance modeling accuracy and data efficiency by leveraging the correlations arising from shared mobility trends, usage behaviors, or regional events. A shared basic model can be pre-trained on aggregated data from multiple sites to learn common traffic dynamics and then fine-tuned to local conditions such as grid-level hotspots or service heterogeneity. Leveraging federated or meta-learning, this strategy enhances robustness in sparsely measured areas, improves data efficiency, and reduces per-site computational cost, enabling faster and more scalable DTM adaptation.
{\bf 3) Autonomous Update of PEM:} 
While changing slowly, large-scale channel statistics and data traffic still evolve over time due to traffic fluctuations and environmental changes (e.g., new obstacles or seasonal variations). Therefore, PEM requires autonomous update mechanisms to adapt to the environment dynamics. Classical adaptive filtering methods can be employed to track and smooth temporal variations in channel and traffic statistics, while online learning can help continuously refine PEM with new data to achieve fast adaptation.

\subsection{Beyond the Physical Layer: Versatile Applications of PEM}

Beyond PHY-layer applications, PEM enables versatile communication designs in higher network layers. At the MAC layer, PEM-captured spatial-temporal traffic statistics allow BSs to proactively schedule users and allocate PRBs to grids with anticipated demand surges. Concurrently, PEM-derived channel knowledge provides statistical cues on link reliability under typical conditions, informing hybrid automatic repeat request (HARQ) parameter adjustments such as retransmission intervals and modulation fallback strategies. At the network layer, PEM supports network slicing design through fine-grained traffic distribution estimates, guiding slice provisioning and RAN-core traffic offloading decisions. The spatial–temporal traffic distribution captured further enhances inter-cell coordination through dynamic handover threshold adjustment and load-aware cell reselection. At the application layer, the predictive traffic knowledge enables QoS-aware service orchestration through content pre-caching in high-demand areas and dynamically adjusting traffic steering rules. By enhancing environmental awareness, PEM supports the development of intelligent, agile and proactive communication designs, thereby improving the adaptability, efficiency and responsiveness of future mobile networks.

\section{Conclusion}

In this article, we have introduced PEMNet, a framework that captures environment knowledge to enable efficient communication in mobile networks. At its core, PEMNet constructs PEMs that capture both large-scale channel statistics and grid-level spatial-temporal traffic patterns using standard-compliant measurements. This design strikes a balance between fidelity and practicality, while jointly embedding channel and traffic information for a richer, more actionable view of the environment. Case studies on multi-cell beamforming and receive beam design showcased the potential of PEMNet-enabled optimization.

This article explores enhancing environment awareness under practical measurement constraints and communication design requirements, aiming to inspire lightweight and high-applicability solutions for future mobile networks. Looking ahead, directions such as multi-modal data fusion, collaborative construction, and autonomous updates merit further study. By empowering versatile designs across different layers, PEMNet advances the realization of autonomous and enhanced environment-aware mobile networks.

\section{Acknowledgments}

The authors thank Tenghao Cai and Tan Song for their support in the case study simulations.





\footnotesize 

\bibliographystyle{IEEEtran}

\bibliography{ref_PEMNet}

@article{traffic_acmSurvey25,
author = {Aouedi, Ons and Le, Van An and Piamrat, Kandaraj and Ji, Yusheng},
title = {Deep Learning on Network Traffic Prediction: Recent Advances, Analysis, and Future Directions},
year = {2025},
issue_date = {June 2025},
publisher = {Association for Computing Machinery},
address = {New York, NY, USA},
volume = {57},
number = {6},
issn = {0360-0300},
doi = {10.1145/3703447},
journal = {ACM Comput. Surv.},
articleno = {151},
numpages = {37}
}

@article{Xinyu_ArXiv25,
  title={A Measurement Report Data-Driven Framework for Localized Statistical Channel Modeling},
  author={Qin, Xinyu and Xue, Ye and Yan, Qi and Zhang, Shutao and Peng, Bingsheng and Chang, Tsung-Hui},
  journal={arXiv preprint arXiv:2509.19342},
  year={2025}
}

@article{Juntao_ArXiv25,
  title={Learning to Gridize: Segment Physical World by Wireless Communication Channel},
  author={Wang, Juntao and Yin, Feng and Ding, Tian and Chang, Tsung-Hui and Luo, Zhi-Quan and Yan, Qi},
  journal={arXiv preprint arXiv:2507.15386},
  year={2025}
}

@article{Binsheng_ArXiv25,
  title={{RF-LSCM}: Pushing Radiance Fields to Multi-Domain Localized Statistical Channel Modeling for Cellular Network Optimization},
  author={Peng, Bingsheng and Zhang, Shutao and Zheng, Xi and Xue, Ye and Qin, Xinyu and Chang, Tsung-Hui},
  journal={arXiv preprint arXiv:2509.13686},
  year={2025}
}

@ARTICLE{SRCON23,
  author={Luo, Zhi-Quan and Zheng, Xi and López-Pérez, David and Yan, Qi and Chen, Xin and Wang, Nanbin and Shi, Qingjiang and Chang, Tsung-Hui and Garcia-Rodriguez, Adrian},
  journal={IEEE Commun. Mag.}, 
  title={{SRCON}: A Data-Driven Network Performance Simulator for Real-World Wireless Networks}, 
  year={2023},
  volume={61},
  number={6},
  pages={96-102},
  doi={10.1109/MCOM.001.2200179}}

@ARTICLE{ERM_WC19,
  author={Bi, Suzhi and Lyu, Jiangbin and Ding, Zhi and Zhang, Rui},
  journal={IEEE Wirel. Commun.}, 
  title={Engineering Radio Maps for Wireless Resource Management}, 
  year={2019},
  volume={26},
  number={2},
  pages={133-141},
  doi={10.1109/MWC.2019.1800146}}

@ARTICLE{Traffic_JSAC19,
  author={Xu, Yue and Yin, Feng and Xu, Wenjun and Lin, Jiaru and Cui, Shuguang},
  journal={IEEE J. Sel. Areas Commun.}, 
  title={Wireless Traffic Prediction With Scalable {Gaussian} Process: {Framework}, Algorithms, and Verification}, 
  year={2019},
  volume={37},
  number={6},
  pages={1291-1306},
  doi={10.1109/JSAC.2019.2904330}}

@ARTICLE{CC_access18,
  author={Studer, Christoph and Medjkouh, SaïD and Gonultaş, Emre and Goldstein, Tom and Tirkkonen, Olav},
  journal={IEEE Access}, 
  title={Channel Charting: Locating Users Within the Radio Environment Using Channel State Information}, 
  year={2018},
  volume={6},
  number={},
  pages={47682-47698},
  doi={10.1109/ACCESS.2018.2866979}
  }

@ARTICLE{CKM_tut24,
  author={Zeng, Yong and Chen, Junting and Xu, Jie and Wu, Di and Xu, Xiaoli and Jin, Shi and Gao, Xiqi and Gesbert, David and Cui, Shuguang and Zhang, Rui},
  journal={IEEE Commun. Surveys Tuts}, 
  title={A Tutorial on Environment-Aware Communications via Channel Knowledge Map for {6G}}, 
  year={2024},
  volume={26},
  number={3},
  pages={1478-1519},
  doi={10.1109/COMST.2024.3364508}
  }

@ARTICLE{REKP_mag25,
  author={Wang, Jialin and Zhang, Jianhua and Zhang, Yuxiang and Sun, Yutong and Nie, Gaofeng and Shi, Lianzheng and Zhang, Ping and Liu, Guangyi},
  journal={IEEE Commun. Mag.}, 
  title={Radio Environment Knowledge Pool for {6G} Digital Twin Channel}, 
  year={2025},
  volume={63},
  number={5},
  pages={158-164},
  doi={10.1109/MCOM.003.2400168}
  }

@ARTICLE{DT_sustainable_mag25,
  author={Xu, Dianlei and Su, Xiang and Tarkoma, Sasu and Hui, Pan},
  journal={IEEE Commun. Mag.}, 
  title={Toward Sustainable {6G} leveraging Digital Twin and Artificial Intelligence: Framework and Case Study}, 
  year={2025},
  volume={},
  number={},
  pages={1-7},
  doi={10.1109/MCOM.003.2400389}
  }

@ARTICLE{LSCM_twc23,
  author={Zhang, Shutao and Ning, Xinzhi and Zheng, Xi and Shi, Qingjiang and Chang, Tsung-Hui and Luo, Zhi-Quan},
  journal={IEEE Trans. Wireless Commun.}, 
  title={A Physics-Based and Data-Driven Approach for Localized Statistical Channel Modeling}, 
  year={2024},
  volume={23},
  number={6},
  pages={5409-5424},
  doi={10.1109/TWC.2023.3326209}}

@article{MR_TS36331,
	title={Evolved Universal Terrestrial Radio Access {(E-UTRA)}; {Radio} Resource Control {(RRC)}; {Protocol} specification (Release 18)},
	author={Technical Specification Group},
	journal={3rd Generation Partnership Project (3GPP), Version 18.4.0, document TS 36.331},
	year={2024}
}

@article{MDT_TS23501,
	title={System architecture for the {5G} System {(5GS)} (Release 19)},
	author={Technical Specification Group},
	journal={3rd Generation Partnership Project (3GPP), Version 19.2.0, document TS 23.501},
	year={2024}
}

@inproceedings{Tenghao_ICASSP24,
  title={Sensing-Assisted Distributed User Scheduling and Beamforming in Muli-Cell {mmWave} Networks},
  author={Cai, Tenghao and Li, Lei and Chang, Tsung-Hui},
  booktitle={Proc. IEEE ICASSP},
  pages={81-85},
  year={2024},
  address = {Seoul, Korea}
}

\section*{Biographies}
\vspace{-1cm}
\begin{IEEEbiographynophoto}{Lei Li}
(lei.ap@outlook.com) is a Postdoc Fellow of CUHK-Shenzhen. His research interests include wireless environment sensing and network optimization.
\end{IEEEbiographynophoto}
\vspace{-0.5cm}
\begin{IEEEbiographynophoto}{Yanqing Xu}
(xuyanqing@cuhk.edu.cn) is a Research Assistant Professor of CUHK-Shenzhen. His research interests include signal processing algorithm designs for massive MIMO systems and pinching-antenna systems. 
\end{IEEEbiographynophoto}
\vspace{-0.5cm}
\begin{IEEEbiographynophoto}{Ye Xue}
(xuey57@mail.sysu.edu.cn) is an Associate Professor of SYSU. Her research spans sparse/efficient AI, physically grounded field world models, and AI for mathematical optimization.

\end{IEEEbiographynophoto}
\vspace{-0.5cm}
\begin{IEEEbiographynophoto}{Feng Yin} 
(yinfeng@cuhk.edu.cn) is an Associate Professor of CUHK-Shenzhen. His research interests include statistical signal processing, Bayesian learning and optimization, and AI empowered sensor fusion applications.
\end{IEEEbiographynophoto}

\vspace{-0.5cm}
\begin{IEEEbiographynophoto}{Chao Shen} 
(chaoshen@sribd.cn) is a Senior Research Scientist of SRIBD, CUHK-Shenzhen. His research interests include large-scale network optimization and integrated sensing and communication.
\end{IEEEbiographynophoto}

\vspace{-0.5cm}
\begin{IEEEbiographynophoto}{Rui Zhang}
[F'17] (elezhang@nus.edu.sg) is a Professor of NUS and also an Adjunct Professor of CUHK-Shenzhen. His research interests include intelligent surfaces, reconfigurable antennas, radio mapping, non-terrestrial communications, wireless power transfer, AI and optimization methods.
\end{IEEEbiographynophoto}

\vspace{-0.5cm}
\begin{IEEEbiographynophoto}{Tsung-Hui Chang}
[F'23] (changtsunghui@cuhk.edu.cn) is a Professor of CUHK-Shenzhen. His research interests include optimization and signal processing methods for wireless communications.
\end{IEEEbiographynophoto}

\ifCLASSOPTIONcaptionsoff
\newpage
\fi

\end{document}